\documentclass[sigconf]{acmart}

\AtBeginDocument{%
  }

\copyrightyear{2026}
\acmYear{2026}
\setcopyright{cc}
\setcctype{by}
\acmConference[ICSE-FoSE]{2026 IEEE/ACM 48th International Conference on Software Engineering: Future of Software Engineering }{April 12--18, 2026}{Rio de Janeiro, Brazil}
\acmBooktitle{2026 IEEE/ACM 48th International Conference on Software Engineering: Future of Software Engineering (ICSE-FoSE), April 12--18, 2026, Rio de Janeiro, Brazil}
\acmPrice{}
\acmDOI{10.1145/3793657.3793890}
\acmISBN{979-8-4007-2479-4/2026/04}

\acmConference[ICSE '26]{48th International Conference on Software Engineering}{April 2026}{Rio de Janeiro, Brazil}

\begin{document}

\title{The Competence Crisis: A Design Fiction on AI-Assisted Research in Software Engineering}

\author{Mairieli Wessel}
\affiliation{%
  \institution{Radboud University}
  \city{Nijmegen}
  \country{The Netherlands}}
\email{mairieli.wessel@ru.nl}

\author{Daniel Feitosa}
\affiliation{%
  \institution{University of Groningen}
  \city{Groningen}
  \country{The Netherlands}}
\email{d.feitosa@rug.nl}

\author{Sangeeth Kochanthara}
\affiliation{%
  \institution{ASTRON}
  \city{Dwingeloo}
  \country{The Netherlands}}
\email{kochanthara@astron.nl}

\begin{abstract}
Rising publication pressure and the routine use of generative AI tools are reshaping how software engineering research is produced, assessed, and taught. While these developments promise efficiency, they also raise concerns about skill degradation, responsibility, and trust in scholarly outputs. This vision paper employs Design Fiction as a methodological lens to examine how such concerns might materialise if current practices persist. 
Drawing on themes reported in a recent community survey, we construct a speculative artifact situated in a near future research setting. The fiction is used as an analytical device rather than a forecast, enabling reflection on how automated assistance might impede domain knowledge competence, verification, and mentoring practices. By presenting an intentionally unsettling scenario, the paper invites discussion on how the software engineering research community  in the future will define proficiency, allocate responsibility, and support learning.
\end{abstract}


\maketitle

\balance

\section{Introduction}

The software engineering research community has grown steadily over the past decade, with flagship conferences receiving increasing numbers of submissions. ICSE submissions rose from 530 in 2016 to 797 in 2023, corresponding to yearly increases of roughly 5 to 15 percent. In the last two years, this growth accelerated sharply, with submissions increasing to 1,049 in 2024, 1,150 in 2025, and 1,469 in 2026, representing increases of around 30 percent year-on-year.\footnote{https://dl.acm.org/conference/icse/proceedings} Other premier software engineering venues including ASE and FSE show similar trend.
Alongside this growth, participation has expanded, and generative AI (genAI) has begun to influence both research topics and conducting research. Despite these trends, dissatisfaction within the community is increasingly visible. Common concerns include a ``peer review crisis'' \cite{future2026}, a gap between academic research and industrial practice, and pressure to prioritize publication counts over depth and long-term relevance. On the other hand, the use of generative AI tools has been shown to accumulate cognitive debt~\cite{kosmyna2025your}.
We note that these are not concerns generated by genAI, but one may argue that they are heavily potentialized by this technological disruption. 

This paper employs Design Fiction \citep{blythe2014research,sterling2009cover} to examine 
the scenario of deep domain knowledge and skill erosion due to extensive genAI usage.
Design Fiction is a research approach that employs narrative artifacts to probe, explore, and critique possible futures of technology~\citep{blythe2014research}. It has been used extensively in human-computer interaction research \citep{encinas2016solution,Muller2018,blythe2016co} and, more recently, within software engineering to reflect on automation, (chat)bots, and future development practices~\citep{wessel2022bots,penney2023anticipating,tan2025revolutionizing}.

We present a fictional story set in a plausible near future. The story does not aim to predict what will happen. Instead, it offers a concrete scenario that extrapolates current pressures and trajectories to the future within the software engineering research community.  By foregrounding social, organizational, and value-related consequences of current trends, the story is intended to support reflection and discussion about the kinds of research culture and practices the community wishes to sustain.

\section{State of the Community}

To assess current sentiment in the software engineering research community and ground the narrative presented in this paper, we analysed themes from open-ended responses to the ICSE 2026 pre-survey~\cite{future2026}. The analysis focused on perceptions of what does and does not work well, sources of stress and joy, and concerns for mentees. Themes were first analysed separately for each question and then synthesised into an overall view of the community’s health. In brief, the findings point to a community at a crossroads.

\par{\textbf{The Bright Side: A Supportive Culture.}}
When asked what works well (Q7), respondents consistently highlighted the community’s inclusive and welcoming culture. Compared to more competitive fields, software engineering is perceived as collegial and welcoming. One respondent remarked, \textit{``Since I was a PhD student I have always felt SE conferences to be a protected and safe environment''} (Q7). In addition, despite acknowledged limitations, the peer review process was often described as more constructive than in neighbouring fields such as Security or AI.

\par{\textbf{Systemic Strains: The Feedback Loop.}}
In contrast, responses to questions on what does not work (Q8) and sources of stress (Q11) reveal a negative feedback loop. Pressure to ``publish or perish'' (Q11) drives an unsustainable volume of submissions, which in turn contributes to what many respondents described as a ``Peer Review Crisis'' (Q8). As one participant stated bluntly, \textit{``Peer review is an utter disaster... reviewing quality is nowadays increasingly low and getting lower; increasingly, acceptance is a lottery''} (Q8).  
These systemic pressures also widen the gap between research and practice, with researchers reporting that they optimise for acceptance rather than for industrial relevance.

\par{\textbf{The Human Cost: Joy vs. Fear.}}
This tension is most apparent when contrasting sources of joy (Q10) with concerns for the future (Q12).
For many respondents, mentorship is the primary source of joy, as illustrated by one comment: \textit{``Supervising graduate students; sharing in their successes, being a good mentor...''} (Q10).
While, this satisfaction is tempered by strong anxiety for those same students. When asked about concerns for mentees (Q12), mentors expressed worry the erosion of deep skills due to over-reliance on Gen AI. One respondent cautioned: \textit{``All the AI-slop that's crowding out more meaningful research is going to make it harder for them...''} (Q12).

This concern, that the next generation may become proficient tool users while losing the ability to engineer from first principles, motivates the design fiction that follows.

\section{The Fiction: \textit{A Review in 2035}}

Dr. Jane Doe was in her office. It was 2035. She was reviewing a manuscript for the Most Prestigious Journal in Software Engineering (MPJSE). The manuscript was titled ``Asynchronous coherence for Bio-Photonic Neural Interfaces.'' It was about a radically new method to connect computers to human brains.

The paper looked very good. The writing was clear and the math was correct. The ``GenAI Contribution Score'' was 98\%, which was normal. But Jane found a problem.

The authors used the Paxos algorithm for the connection. This is a common algorithm for computers to agree on data. But this was the mistake.

Jane checked the math. Biology is not like a computer. Neural signals are continuous and noisy. You cannot use a discrete algorithm like Paxos. It would cause seizures.

She paused and looked out the window. How could they miss this? She opened her own console. She did not use GenAI. She used an old Python script she wrote five years ago. She simulated the interface.

\begin{verbatim}
> import numpy as np
> signal = bio_signal_generator(type='cortex')
> consensus = paxos.run(signal)
> ERROR: Signal discontinuity. 
> PREDICTION: Cascade failure (Seizure).
\end{verbatim}

The simulation confirmed her fear. It took her five minutes to check. Why did the authors not check?

She looked at the authors’ names. They were productive and successful researchers who had been trained to do research with GenAI tools. These tools were reliable and served as very competent research assistants. The authors knew how to write prompts and let the AI handle the details.

The AI model made a mistake. It saw ``distributed nodes'' in the problem description and used Paxos from its training data. It did not understand the biology. The authors did not understand it either. They accepted the AI's answer. They might not have the domain knowledge or the skill to check if it was correct.

Jane felt cold. These researchers did not know the basics of the domain. If the AI made a mistake on a new problem, they could not fix it. They were not engineers anymore. They were just users!

It was not just an isolated incident. Jane realized with a sinking feeling that this was the third paper she reviewed this month with a similar ``hallucination.'' A dangerous pattern was emerging in the burgeoning field of Bio-Photonic Interfaces: a systematic failure where plausible-sounding but physically impossible methods were being accepted as fact. The truly scary part was not this rejection, but the flourishing wave of \emph{published} works already in the digital library. Flawed papers citing other flawed papers, creating a feedback loop that future models would only reinforce. The well of knowledge was being poisoned, and no one was checking the water. \textbf{\emph{By failing to self-correct, the trust of industry (and society) in this research community was in jeopardy.}}

\section{The Rejection (A Call for Competence)}

Jane pulled up the review form. The cursor blinked. She realized she wasn't just writing to these authors; she was writing to the whole community. She typed, deleted, and typed again, trying to articulate a warning that went beyond this single manuscript.\\

\noindent
\textbf{Manuscript:} \#2035-0842\\
\textbf{Recommendation:} Reject

\vspace{0.5em}
\textit{I am recommending rejection for this paper, but the ``Contribution Score'' or the specific technical error is not the primary reason. The reason is the systemic fragility this work represents.}

\textit{Technically, the paper is a seamless execution of a flawed premise. The application of discrete consensus algorithms (Paxos) to continuous biological substrates violates the basic principles of neurophysiology. As I verified via manual simulation, implementing this protocol would result in catastrophic biological failure.}

\textit{However, what is most alarming is that this paper does not stand alone. It cites three other recent papers that make similar ``hallucinated'' assumptions. These cited papers likely passed through the peer-review because they \textit{looked} correct to an AI-assisted reviewer or a non-expert. We are witnessing the poisoning of our own well. We are building a citation graph on a foundation of convincingly-sounding nonsense.}

\textit{This is the ``Competence Crisis'' we feared a decade ago. We have become so adept at operating our tools that we have neglected the deep, `dirty' details of the domains we claim to engineer. We are trusting models to solve problems they have never seen. We lack the fundamental friction (with the raw problem) required to verify them.}

\textit{I recommend the rejection of this paper not only because it is built on fundamentally flawed premise but also contributes to a feedback loop of misinformation. If we cannot reason from first principles, if we cannot validate the output of our tools against grounded checks, we are no longer engineers. We are merely operators of black boxes.}

\textit{I urge the authors: do not just `fix' the algorithm. Go back to the bench. Re-derive the physics. Break the cycle of recursive generation. We need you to understand the problem, not just prompt.}

\section{Discussion: From Fiction to Action}

Jane's story is a fiction, but the threat it illustrates is the `scary part' of our community's current trajectory. The ICSE 2026 pre-survey~\cite{future2026} highlighted a fear of the erosion of fundamental research skills. In our fiction, this erosion has consequences far graver than a bad paper.

\subsection{The Loss of Technical Grounding}
The GenAI model failed because it encountered an out-of-distribution problem. There is a new context (bio-photonics) for which its training data was either misleading or not present. This is the inevitability of research; we exist to explore the unknown. If our primary method of inquiry is to ask a model trained on the \textit{known}, we inherently limit our capacity to true innovation.

We know this problem well when it comes to the Software Engineering profession. Ideally, professionals have a `T-shaped' profile. This also applies to us researchers. We (should) strive to be T-shaped: possessing a broad understanding of the field (the horizontal bar) and deep, specialized expertise in a specific domain (the vertical bar). GenAI acts as a powerful lever for the horizontal bar, allowing researchers to instantly summarize literature and connect concepts across disciplines. However, by removing the friction of learning, e.g., the hours of debugging, deriving, and failing, we are causing an athrophy of the vertical bar.

The danger is that we are cultivating a generation of \textbf{`dash-shaped'} researchers: professionals with immense breadth and surface-level fluency, but with no deep vertical grounding. In our fiction, the authors were `dashes'. They could connect Paxos to Biology because the genAI tool made it easy, but they lacked the vertical depth to know why that connection was physically impossible. Jane, by contrast, retained her `T-shape,' giving her the depth required to verify the specifics. If we rely on genAI to bridge the gap between problem and solution, who will have the technical depth to build the bridge manually when the model fails?

\subsection{Value-Sensitive Design for our Community}

To reason about the challenges illustrated in the fiction, we draw on Value Sensitive Design (VSD) \cite{friedman1996value,friedman2019value}. VSD is a research and design approach that explicitly accounts for human values throughout the design, development, and use of technology. Rather than treating values as secondary concerns or downstream effects, VSD treats them as integral to technical decisions, asking whose values are embedded in a system, who benefits from them, and who bears the costs.

Applying VSD to the software engineering research community shifts attention from individual tools to the socio-technical system in which those tools are used. The story highlights how genAI-assisted workflows can subtly reshape researchers' roles, responsibilities, and learning practices. When automated systems handle large parts of technical reasoning, researchers may lose opportunities to exercise judgment and to develop or maintain foundational skills. This has implications for \textbf{human agency}, as researchers risk becoming dependent on tools rather than remaining active decision-makers in the research process.

VSD also puts forward questions of \textbf{accountability}. In the fiction, the error is not simply a technical failure but a failure of responsibility, as neither the authors nor their reviewers identify a flawed assumption. When genAI systems are deeply embedded in research workflows, it becomes unclear where responsibility lies when results are wrong. VSD argues that systems should be designed and adopted in ways that keep humans accountable for outcomes, rather than allowing responsibility to be implicitly shifted onto tools.

Finally, the community survey highlights \textbf{mentorship} as both a source of satisfaction and a source of concern. From a VSD perspective, mentorship is not only a social practice but also a value that can be supported or undermined by technology. If research workflows minimize struggle, experimentation, and failure, they may also reduce opportunities for learning and guidance. Applying VSD encourages the community to reflect on how genAI tools affect how researchers are trained, how expertise is developed, and how knowledge is passed on.

Viewed through a VSD lens, the fiction is not a warning against genAI use itself, but against unreflective adoption that neglects the values the community claims to hold. It suggests that choices about tools, incentives, and evaluation practices should be made with explicit attention to agency, responsibility, and learning, rather than efficiency alone.


\subsection{Action Plan}

This narrative calls for a rethinking of how research and education are organised in an environment where AI systems are part of everyday practice. A first priority is \textbf{\textit{renewed emphasis on first-principles thinking}}, where proposed solutions are evaluated against fundamental properties of the problem domain rather than accepted because they match familiar patterns or appear internally consistent. Such reasoning remains essential when working at the boundaries of existing knowledge, where prior data and learned correlations are unreliable.

A second priority is \textbf{\textit{verification literacy}}. Researchers must retain the ability to audit and test AI-generated outputs in a rigorous way. This requires deeper technical understanding, not less, since meaningful verification depends on knowing what assumptions are being made and where they may break down. Without this capability, errors risk propagating unchecked through review and follow-up work.

Finally, the community should support \textbf{\textit{resilience in research practice}}, understood as the ability to reason about and solve problems without continuous reliance on automated systems. Maintaining this capacity may require intentional friction in both education and practice, for example by designing learning activities that slow down interaction with tools and force explicit reasoning \cite{chen2024exploring}. Such approaches help preserve core engineering skills that are difficult to recover once lost.

As AI-based tools become more capable and more widespread, the challenge is not to avoid their use, but to ensure that researchers remain responsible for their outputs and capable of intervening when these tools fail.



\section{Conclusion}
Our fictional review highlights how errors become dangerous not when tools fail, but when human oversight erodes. When researchers are unable or unwilling to interrogate assumptions, validate outputs, or step outside automated pipelines, mistakes can propagate through citation networks and review processes with little resistance. In such settings, competence becomes difficult to assess, accountability becomes diffuse, and trust in the research record is put at risk.

Addressing these risks requires attention to how the community organises learning, evaluation, and mentorship under conditions of widespread AI assistance. Practices that preserve technical grounding, encourage verification, and maintain human responsibility are not incidental, they shape what kinds of researchers the community produces. The fiction offered here is intended as a prompt for collective reflection on these choices, and on how the software engineering research community wishes to define competence in an AI mediated future.

\bibliographystyle{ACM-Reference-Format}
\bibliography{main}

\end{document}